\documentclass[twocolumn]{aastex62}

\usepackage{textcomp}
\usepackage{bm,color}
\usepackage{verbatim}
\usepackage{amsmath}
\usepackage{ulem}
\usepackage{hyperref}  
\usepackage{booktabs}
\usepackage{bigstrut}
\usepackage{xspace}
 \usepackage{multirow}

\newcommand{\tabref}{Table\xspace}

\newcommand{\pb}{\textsc{Polarbear}\xspace}
\newcommand{\sa}{Simons Array}

\newcommand{\lcdm}{$\Lambda$CDM\xspace}

\newcommand{\nver}{\hat{\mathbf{n}}}
\newcommand*\Bell{\ensuremath{\boldsymbol\ell}}

\newcommand{\sqdeg}{$\deg^2$}

\newcommand{\beq}{\begin{equation}}
\newcommand{\eeq}{\end{equation}}
\newcommand{\bea}{\begin{eqnarray}}
\newcommand{\eea}{\end{eqnarray}}

\newcommand{\signif}{4.8}
\newcommand{\signifPearson}{4.3}
\newcommand{\bAvalue}{\ensuremath{5.69 \pm 1.30}}
\newcommand{\bAvaluePlanck}{\ensuremath{3.40 \pm 0.51}}

\newcommand{\bvalue}{\ensuremath{5.76\pm 1.25}}
\newcommand{\bvaluePlanck}{\ensuremath{3.43\pm 0.51}}
\newcommand{\bvaluePearson}{\ensuremath{5.38\pm 1.40}}
\newcommand{\Mhalovalue}{\ensuremath{13.5^{+0.2}_{-0.3}}}
\newcommand{\MhalovaluePlanck}{\ensuremath{12.9^{+0.2}_{-0.2}}}

\newcommand{\APC}{AstroParticule et Cosmologie, Univ Paris Diderot, CNRS/IN2P3, CEA/Irfu, Obs de Paris, Sorbonne Paris Cit\'e, France}

\newcommand{\damtp}{Department of Applied Mathematics and Theoretical Physics, University of Cambridge, Cambridge CB3 0WA, UK}
\newcommand{\DFI}{Departamento de F\'isica, FCFM, Universidad de Chile, Blanco Encalada 2008, Santiago, Chile}

\newcommand{\IFPU}{Institute for Fundamental Physics of the universe (IFPU), Via Beirut 2, 34151, Grignano (TS), Italy}
\newcommand{\INFN}{The National Institute for Nuclear Physics, INFN, Sezione di Trieste Via Valerio 2, I-34127, Trieste, Italy}
\newcommand{\IPMU}{Kavli IPMU (WPI), UTIAS, The University of Tokyo, Kashiwa, Chiba 277-8583, Japan}
\newcommand{\IpmuBerkeley}{Kavli Institute for the Physics and Mathematics of the universe (WPI), Berkeley Satellite, the University of California, Berkeley 94720 , USA}  
\newcommand{\JAXA}{Institute of Space and Astronautical Science, Japan Aerospace Exploration Agency (JAXA), Sagamihara, Kanagawa 252-0222, Japan}
\newcommand{\JHU}{Department of Physics and Astronomy, Johns Hopkins University, Baltimore, MD 21218, USA}
\newcommand{\KEK}{Institute of Particle and Nuclear Studies, High Energy Accelerator Research Organization (KEK), Tsukuba, Ibaraki 305-0801, Japan}
\newcommand{\Kyoto}{Department of Physics, Kyoto University, Kyoto, Kyoto 606-8502, Japan}
\newcommand{\LBL}{Physics Division, Lawrence Berkeley National Laboratory, Berkeley, CA 94720, USA}

\newcommand{\Melbourne}{School of Physics, University of Melbourne, Parkville, VIC 3010, Australia}

\newcommand{\PUCC}{Centro de Astro-Ingenier\'ia, Pontificia Universidad Cat\'olica de Chile, Vicu\~na Mackenna 4860, Santiago, Chile}
\newcommand{\RAL}{Radio Astronomy Laboratory, University of California Berkeley, Berkeley, CA 94720, USA}
\newcommand{\RESCEU}{Research Center for the Early universe, School of Science, The University of Tokyo, Tokyo 113-0033, Japan}
\newcommand{\SISSA}{International School for Advanced Studies (SISSA), Via Bonomea 265, 34136, Trieste, Italy}
\newcommand{\SOKENDAI}{Department of Particle and Nuclear Physics, SOKENDAI, Hayama, Kanagawa 240-0193, Japan}
\newcommand{\SSL}{Space Sciences Laboratory, University of California, Berkeley, CA 94720, USA}
\newcommand{\Stanford}{Department of Physics and KIPAC, Stanford University, Stanford, CA 94305, USA}
\newcommand{\Sussex}{Department of Physics \& Astronomy, University of Sussex, Brighton BN1 9QH, UK}

\newcommand{\Tokyo}{Department of Physics, The University of Tokyo, Bunkyo-ku, Tokyo 113-0033, Japan}

\newcommand{\UCBP}{Department of Physics, University of California Berkeley, Berkeley, CA 94720, USA}

\newcommand{\UCSD}{Department of Physics, University of California, San Diego, CA 92093-0424, USA}
\newcommand{\UNM}{Department of Physics and Astronomy, University of New Mexico, Albuquerque, NM 87131,  USA}

\newcommand{\UrbanaC}{Department of Physics, University of Illinois at Urbana-Champaign, Urbana, Illinois 61801, USA}

\begin{document}
\title{Cross-correlation of  \pb CMB Polarization Lensing with High-$z$ Sub-mm \textit{Herschel}-ATLAS Galaxies}

\author[0000-0002-1571-663X]{M. Aguilar Fa\'undez}
\affiliation{\JHU}
\affiliation{\DFI}

\author[0000-0002-3407-5305]{K. Arnold}
\affiliation{\UCSD}

\author[0000-0002-8211-1630]{C. Baccigalupi}
\affiliation{\SISSA}
\affiliation{\IFPU}
\affiliation{\INFN}

\author[0000-0002-1623-5651]{D. Barron}
\affiliation{\UNM}

\author[0000-0003-0848-2756]{D. Beck}
\affiliation{\APC}

\author[0000-0003-4847-3483]{F. Bianchini}
\affiliation{\Melbourne}

\author{D. Boettger}
\affiliation{\PUCC}

\author{J. Borrill}
\affiliation{\LBL}
\affiliation{\SSL}

\author[0000-0002-5751-1392]{J. Carron}
\affiliation{\Sussex}

\author[0000-0002-7764-378X]{K. Cheung}
\affiliation{\UCBP}

\author[0000-0002-3266-857X]{Y. Chinone}
\affiliation{\UCBP}
\affiliation{\IPMU}

\author[0000-0001-5471-3434]{H. El Bouhargani}
\affiliation{\APC}

\author{T. Elleflot}
\affiliation{\UCSD}

\author{J. Errard}
\affiliation{\APC}

\author[0000-0002-3255-4695]{G. Fabbian}
\affiliation{\Sussex}

\author{C. Feng}
\affiliation{\UrbanaC}

\author{N. Galitzki}
\affiliation{\UCSD}

\author{N. Goeckner-Wald}
\affiliation{\UCBP}

\author[0000-0003-1443-1082]{M. Hasegawa}
\affiliation{\KEK}

\author{M. Hazumi}
\affiliation{\KEK}
\affiliation{\IPMU}
\affiliation{\SOKENDAI}
\affiliation{\JAXA}

\author{L. Howe}
\affiliation{\UCSD}

\author{D. Kaneko}
\affiliation{\IPMU}

\author{N. Katayama }
\affiliation{\IPMU}

\author[0000-0003-3118-5514]{B. Keating }
\affiliation{\UCSD}

\author{N. Krachmalnicoff}
\affiliation{\SISSA}

\author{A. Kusaka}
\affiliation{\LBL}
\affiliation{\Tokyo}
\affiliation{\IpmuBerkeley}
\affiliation{\RESCEU}

\author{A.~T.~Lee}
\affiliation{\UCBP}
\affiliation{\LBL}
\affiliation{\RAL}

\author{D. Leon}
\affiliation{\UCSD}

\author{E. Linder}
\affiliation{\LBL}

\author{L.~N.~Lowry}
\affiliation{\UCSD}

\author[0000-0003-0041-6447]{F. Matsuda}
\affiliation{\IPMU}

\author[0000-0003-2176-8089]{Y. Minami}
\affiliation{\KEK}

\author[0000-0001-8366-3755]{M. Navaroli}
\affiliation{\UCSD}

\author[0000-0003-0738-3369]{H. Nishino}
\affiliation{\KEK}

\author{A.~T.~P.~Pham}
\affiliation{\Melbourne}

\author[0000-0001-9807-3758]{D. Poletti}
\affiliation{\SISSA}
\affiliation{\INFN}

\author[0000-0002-0689-4290]{G. Puglisi }
\affiliation{\Stanford}

\author{C.~L.~Reichardt}
\affiliation{\Melbourne}

\author{B.~D.~Sherwin}
\affiliation{\damtp}

\author[0000-0001-7480-4341]{M. Silva-Feaver}
\affiliation{\UCSD}

\author[0000-0002-9777-3813]{R. Stompor}
\affiliation{\APC}

\author[0000-0001-8101-468X]{A. Suzuki}
\affiliation{\LBL}

\author{O. Tajima}
\affiliation{\Kyoto}

\author[0000-0001-9461-7519]{S. Takakura}
\affiliation{\IPMU}

\author{S. Takatori}
\affiliation{\SOKENDAI}
\affiliation{\KEK}

\author{G. P.~Teply}
\affiliation{\UCSD}

\author{C. Tsai}
\affiliation{\UCSD}

\author[0000-0002-3942-1609]{C. Verg\`{e}s}
\affiliation{\APC}

\collaboration{The \pb Collaboration}
\noaffiliation

\correspondingauthor{F. Bianchini}

\begin{abstract}

We report a  \signif$\sigma$ measurement of the cross-correlation signal between the cosmic microwave background (CMB) lensing convergence reconstructed from measurements of the CMB polarization made by the \pb experiment and the infrared-selected galaxies of the \textit{Herschel}-ATLAS survey. This is the first measurement of its kind. We infer a best-fit galaxy bias of $b = \bvalue$, corresponding to a host halo mass of $\log_{10}(M_h/M_\odot) =\Mhalovalue$ at an effective redshift of  $z \sim 2$ from the cross-correlation power spectrum. 
Residual uncertainties in the redshift distribution of the sub-mm galaxies are subdominant with respect to the statistical precision. 
We perform a suite of systematic tests, finding that instrumental and astrophysical contaminations are small compared to the statistical error.
This cross-correlation measurement only relies on CMB polarization information that, differently from CMB temperature maps, is less contaminated by galactic and extragalactic foregrounds, providing a clearer view of the projected matter distribution. This result demonstrates the feasibility and robustness of this approach for future high-sensitivity CMB polarization experiments.
\end{abstract}

\keywords{cosmology: observations, cosmology: cosmic background radiation, cosmology: large-scale structure of the universe, galaxies: high-redshift}


\section{Introduction}
The pattern of the cosmic microwave background (CMB) anisotropies not only provides a snapshot of the primordial universe at $z \approx 1100$, but also encodes a wealth of information about its evolution after recombination \citep[e.g.,][]{Aghanim2008}. 
In particular, the trajectory of CMB photons while traveling between the last-scattering surface and us can be deflected by the intervening matter distribution, an effect known as weak gravitational lensing \citep{Lewis2006}. 
These deflections, typically of a few arcminutes, 
introduce correlations between modes of the CMB anisotropies that can be exploited to reconstruct the projected gravitational potential (CMB lensing potential $\phi$) in the whole observable universe \citep{Hu2002,Hirata2003}. 
The sensitivity of the lensing signature to both the geometry and the growth of structures of the universe makes it suitable to break the geometrical degeneracy affecting the primary CMB \citep{Stompor1999} and to investigate the neutrino and dark sector. 

Since its first detection about a decade ago \citep{Smith2007,Kuo2007,Hirata2008}, CMB lensing science has rapidly progressed and several collaborations have reported highly significant measurements of the CMB lensing power spectrum, including the ACT \citep[][temperature and polarization]{Das2014,Sherwin2017}, BICEP/\textit{Keck} \citep[][polarization-only]{BICEP22016}, \textit{Planck} \citep[][temperature and polarization]{Ade2014c,PlanckCollaboration2015,PlanckLensing2018}, \pb \citep[][polarization-only]{POLARBEARLensing}, and SPT \citep[][temperature and polarization]{Story2015,Omori2017} collaborations. 

Given that CMB lensing probes the projected matter distribution along the line-of-sight up to very high redshifts, it is highly correlated with other tracers of large-scale structure (LSS) such as galaxies.
Several groups have detected the cross-correlation signal between CMB lensing and galaxies selected in different wavelengths. 
Cosmological and astrophysical applications of the CMB lensing-galaxy clustering cross-correlations include the study of the galaxy bias evolution \citep[e.g.,][]{Sherwin2012,Bleem2012,DiPompeo2014,Bianchini2015,Allison2015,Bianchini2016}, the measurement of the growth of structure \citep[e.g.,][]{Giannantonio2016,Pullen2016,Bianchini2018,Peacock2018,Omori2018}, the calibration of cosmic shear measurements \citep{Baxter2016}, and the investigation of primordial non-Gaussianities \citep{Giannantonio2014}. 
Moreover, cross-correlations are becoming a standard probe to be included in the general cosmological parameters estimation framework \citep[e.g.,][]{Abbott2018}.
The advantage of a cross-correlation analysis is twofold. 
First, cross-correlation allows one to separate the CMB lensing signal to a specific range of redshifts (the redshifts of the galaxy sample). 
Second, cross-correlations are less prone to systematic effects as most systematics will be uncorrelated between different experiments and wavelengths.

In this paper, we measure the cross-correlation between CMB lensing convergence maps $\kappa = -\frac{1}{2}\nabla^2\phi$ reconstructed by the \pb experiment and the clustering of bright sub-millimetre (sub-mm) galaxies detected by the \textit{Herschel} satellite. 
Sub-mm galaxies are thought to undergo an intense phase of star-formation in a dust-rich environment, where ultraviolet light emitted by newly born stars is absorbed by the dust and subsequently re-emitted in the far-infrared \citep[e.g.,][]{Smail1997,Blain2002}. The brightest sub-mm galaxies can reach luminosities of about $10^{13}L_{\odot}$, with corresponding star-formation rates up to $\sim 1000 \, M_{\odot}/$year. 

A peculiarity of the spectral energy distribution (SED) of sub-mm galaxies is that there is a strongly negative $K$-correction at mm and sub-mm wavelengths, meaning that the observed sub-mm flux of such galaxies is nearly independent of redshift from $1 \lesssim z \lesssim 8$  \cite[for a recent review of dusty star-forming galaxies see][]{Casey2014}. 
Sub-mm galaxy samples are weighted towards high redshifts ($1 \lesssim z \lesssim 3$), which is the redshift range where a given matter fluctuation will lead to the largest CMB lensing signal. 
Thus sub-mm galaxies are perfect candidates for CMB lensing-galaxy density cross-correlation studies.

The Cosmic Infrared Background (CIB) is thought to comprise the emission of unresolved infrared galaxies. It is then natural to expect a high degree of correlation with CMB lensing \citep{Song2003}. Recent studies have investigated the cross-correlation between CMB lensing and maps of the diffuse CIB \citep[e.g.][]{Holder2013,Ade2014a,PBIPPxHerschel,Engelen2015, PlanckLensing2018}, finding correlation coefficients up to 80\% at about 500 $\mu$m. 

Our analysis clearly shares some similarities with these studies because \textit{Herschel} galaxies constitute part of the CIB, even though, despite having been extensively studied in recent years, the exact redshift distribution of contributions to the CIB is still debated \citep[e.g.,][and references therein]{Casey2014}. CIB only provides an integrated measurement and thus, unlike the catalog-based approach adopted in this work, prevent any accurate redshift tomography of the cross-correlation signal to study the properties of sub-mm galaxies.

In the past, \citet{Bianchini2015,Bianchini2016} cross-correlated similar sub-mm \textit{Herschel} catalogs with the 2013 and 2015 \textit{Planck} CMB lensing maps, reporting a rejection of the no-correlation hypothesis between the two fields at the $\approx 20\sigma$ level. While \textit{Planck} CMB lensing reconstruction is mostly dominated by the CMB temperature information, and hence more contaminated by galactic and extragalactic foregrounds, \pb CMB lensing convergence maps only rely on polarization data. This represents the first study of this kind and was made possible thanks to the depth of the \pb observations. With this sensitivity and \pb resolution, probes of cross-correlation signal at smaller scales have become accessible.

\pb map depth sensitivity is comparable to upcoming ground-based CMB experiments which will cover a much larger fraction of the sky. This work thus serves as a proof of concept that reliable cross-correlation measurements can indeed be achieved without CMB temperature information, providing a more robust measurement against galactic and extragalactic foregrounds \citep[e.g.][]{Smith2008,Challinor2017}.

The outline of this paper is as follows: in Sec.~\ref{sec:data} we introduce the datasets used in this analysis and in Sec.~\ref{sec:theory} we briefly review the theoretical background of CMB lensing-galaxy cross-correlation. The analysis methods are described in Sec.~\ref{sec:methods}, while results are presented in Sec.~\ref{sec:results}. Finally, we draw our conclusions in Sec.~\ref{sec:conclusions}. 

Throughout the paper, unless otherwise stated, we assume a flat \lcdm cosmological model described by the best-fit parameters from \textit{Planck} 2018 TTTEEE + lowE + lensing chains provided by \citet{PlanckParameters2018}. 

\section{Data and Simulations}
This work cross-correlates CMB lensing maps from the \pb experiment and a galaxy overdensity field from \textit{Herschel}-ATLAS. 
In this section, we describe both datasets, as well as the simulations used to construct the employed statistical estimators and error bar estimation.
\label{sec:data}
\subsection{\pb Convergence Map}
The \pb experiment consists of an array of 1274 polarization-sensitive transition-edge sensors observing in a spectral band centered at 148 GHz installed on the 2.5 m primary aperture Huan Tran Telescope at the James Ax Observatory in Chile \citep{Arnold2012, Kermish2012}.  
The lensing convergence map used in this work has been reconstructed using the $Q$ and $U$ Stokes parameters maps of the first two observing seasons, from May 2012 to June 2013 and from October 2013 to April 2014 \citep{POLARBEAR2017} (hereafter PB17).
Among the three fields observed during this period, we used those overlapping with the \textit{Herschel}-ATLAS survey, which are centered at (RA,Dec) = (23$^{\rm h}$12$^{\rm m}$14$^{\rm s}$, -32$^{\circ}$48') and (11$^{\rm h}$53$^{\rm m}$0$^{\rm s}$,-0$^{\circ}$30'), and will be referred to as RA23 and RA12, respectively, in the following. 
Each field encompasses  a sky area of roughly 10 deg$^2$ with polarization noise levels of 5 and 6 $\mu$K-arcmin.

The lensing reconstruction procedure adopted the quadratic estimator algorithm by \citet{Hu2002}. For each field, 
we construct an apodized mask from the smoothed inverse-variance weights of the \pb map after masking out the 
pixels that are within 3 arcmin of radio sources contained in the ATCA catalog (we find five and four sources in RA12 
and RA23 respectively). 
The input of the lensing quadratic estimator is a set of optimally filtered $E$ and $B$ harmonic coefficients. Similar to \citet{Story2015,PlanckLensing2018}, we Wiener-filter the input  $Q$ and $U$ maps to down-weight noise-dominated modes, as well as to deconvolve for the transfer function, beam, pixelation, and masking effects. Assuming that the data maps are composed by the sum of a sky signal, a sky noise, and pixel domain noise, we perform an inverse-variance filtering and output the $E/B$ multipoles. Note that only $E/B$ harmonic modes between $500 \le \ell \le 2700$ are retained before being passed to the quadratic estimator.
For each patch, we reconstruct the CMB lensing convergence $\kappa$ using the $EE$ and $EB$ estimators as

\beq
\hat{\kappa}^{XY}(\mathbf{L}) =  A_L^{XY} \int \frac{d^2\Bell}{(2\pi)^2} w^{XY}_{\Bell,\mathbf{L}}\bar{X}(\Bell)\bar{Y}
(\mathbf{L}-\Bell),
\eeq
where $\bar{X}$ and $\bar{Y}$ are either the filtered $E$- or $B$-modes, $\Bell$ and $\mathbf{L}$ are the 
wavevectors in the two-dimensional Fourier space, $A_L^{XY}$ is a function that normalizes the estimate, and 
$w^{XY}_{\Bell,\mathbf{L}}$ denotes the lensing weight functions (see \citet{Hu2002} for the exact expressions). 
More details about the reconstruction of CMB lensing with \pb can be found in \citet{davidspaper}. 

\subsection{Lensing Convergence Maps Simulations}
Simulated reconstructed convergence maps are used to normalize the quadratic lensing estimator as well as to estimate a mean-field map $\langle \hat{\kappa} \rangle$ that is subtracted from the reconstructed \pb lensing map. This mean-field map takes into account the statistical anisotropy induced by masking and inhomogeneous noise that introduces a spurious statistical anisotropy that affects the quadratic estimator. We produce a single simulated lensing convergence map by generating a Gaussian realization of an unlensed CMB field that we remap in the pixel domain according to a deflection field computed with the Born approximation \citep{fabbian2013, lewis2005}. The deflection field is computed as the gradient of a Gaussian realization of the CMB lensing potential that includes the non-linear corrections to its variance predicted using the \texttt{Halofit} prescription \citep{halofit,Takahashi2012}. Realistic noise is added to the simulated signal time-ordered-data (TOD) that is created by scanning the noiseless lensed CMB map. The TODs are then mapped using the PB17 pipeline A mapmaking algorithm and later processed through the lensing estimation pipeline. The pipeline A mapmaking algorithm is based on the MASTER method \citep{Hivon2002} and we refer the reader to PB17 for further details. The simulation procedure neglects the non-Gaussianity of the matter distribution induced by non-linear gravitational collapse and post-Born effects that could bias lensing estimators \citep{pratten2016, bohm2016, beck2018,Bohm2018} as these effects are negligible at \pb sensitivities. The lensing convergence simulations are also used for the band powers covariance estimation on the final cross-power spectrum measurement as discussed in Sec.~\ref{sec:methods}. 

\subsection{Galaxy Overdensities Map}
\label{sec:hatlas}
We used publicly available data\footnote{Available at \url{http://www.h-atlas.org/public-data/download}.} from the \textit{Herschel} Astrophysical Terahertz Large Area Survey \citep[H-ATLAS,][]{Eales2010a}. 
H-ATLAS is an open-time key program on the \textit{Herschel} Space Observatory \citep{Pilbratt2010}  that has surveyed about 600 deg$^2$ of the sky in five bands between 100 and 500 $\mu$m with two cameras, the Photodetector Array Camera and Spectrometer \citep[PACS,][]{Poglitsch2010} and the Spectral and Photometric Imaging Receiver \citep[SPIRE,][]{Griffin2010}. 
Two of the survey's five fields overlap with the \pb survey -- the South Galactic Pole (SGP) and GAMA 12 (G12) fields. 
The H-ATLAS mapmaking is described by \citet{Valiante2016,Smith2017}, while the source extraction, catalogue generation, and optical identification can be found in \citet{Bourne2016,Maddox2018}.

\begin{figure}
	\includegraphics[width=\columnwidth]{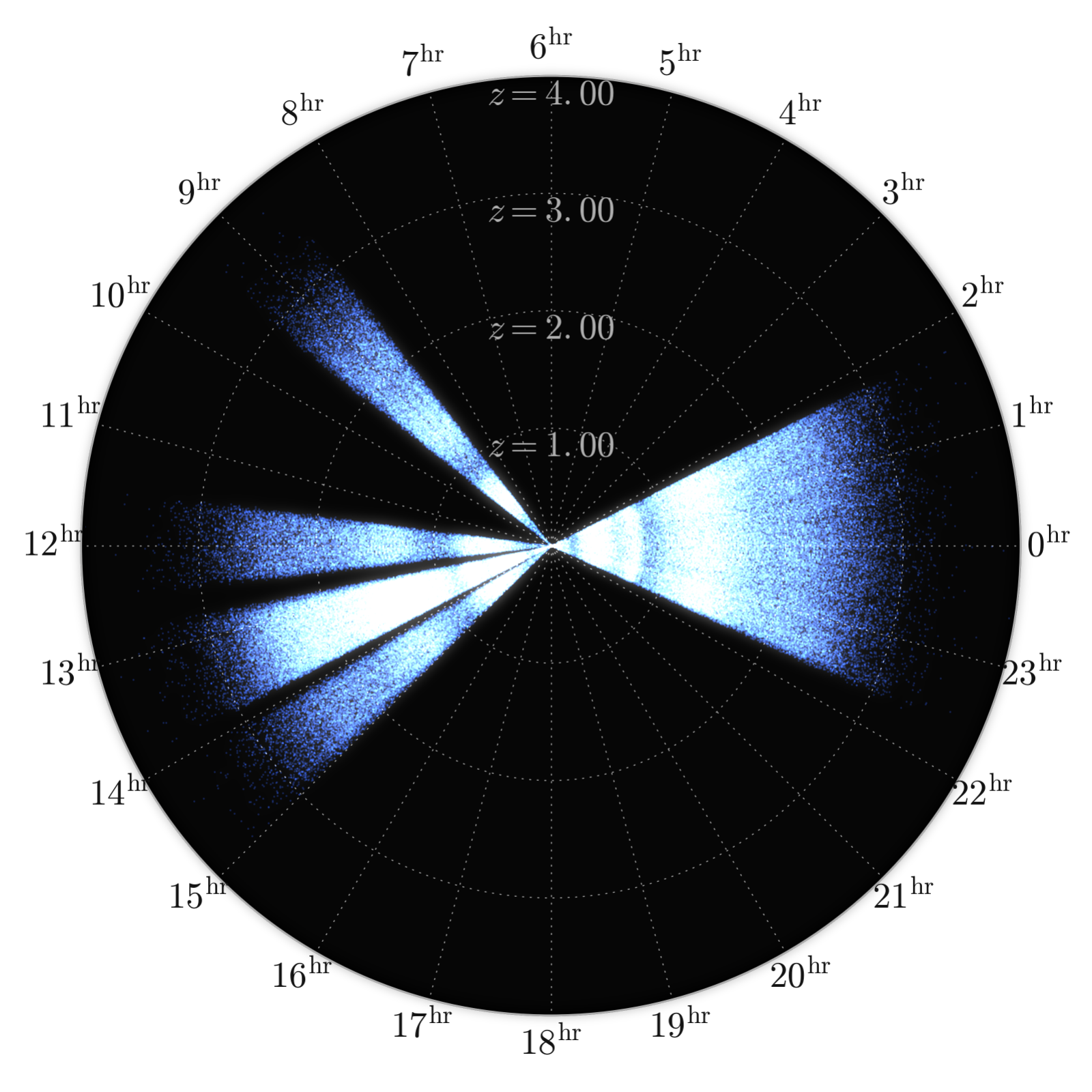}
    \caption{The radial and angular distribution of sub-mm galaxies in all of the five \textit{Herschel}-ATLAS patches. Note that photometric redshifts have been used to place sources along the redshift axis.}
    \label{fig:H-ATLAS}
\end{figure}

With its 3.5 m primary mirror (the largest one currently in space), \textit{Herschel} represented a huge leap forward in the field of sub-mm/far-IR astronomy since all of its predecessors were severely limited by poor angular resolution, a restricted wavelength observational range, and observations were only available over small patches of the sky. By operating at a diffraction limited resolution over $100\ \mu\text{m} \lesssim \lambda \lesssim 500\ \mu\text{m}$, thus covering most of the dust emission of typical galactic spectral energy distribution (SED), \textit{Herschel} has been able to pierce the distant universe, increasing the number of known sub-mm sources from hundreds to hundreds of thousands.
As can be seen in Fig.~\ref{fig:H-ATLAS}, the sub-mm galaxies detected by H-ATLAS span a wide range of redshifts, from the local universe \citep{Eales2018} up to a redshift of about 6 \citep{Zavala2018}, and can be broadly split in two main populations. 
The low-$z$ ($z \lesssim 1$) population is mostly composed of normal late-type and star-burst galaxies with low to moderate star formation rates (SFRs) \citep{Dunne2011,Guo2011,Amvrosiadis2018} while the high-$z$ galaxies tend to have high SFRs (higher than few hundreds $M_{\odot}$yr$^{-1}$) and are much more strongly clustered \citep{Maddox2010,Xia2012,Amvrosiadis2018}. 
The properties of the high-$z$ population suggest that these sources are the progenitors of local massive elliptical galaxies \citep{Lapi2011,Amvrosiadis2018}.

Following \citet{Bianchini2015,Bianchini2016}, we select the galaxy sample used in this work from the full-sky H-ATLAS catalog adopting selection criteria to isolate the best high-$z$ ($z \gtrsim 1$) tracers of the large-scale structures that contribute to the CMB lensing signal:
\begin{enumerate}
\item flux density at 250 $\mu$m larger than $S_{250 \rm \mu m} > 35$ mJy;
\item $\ge 3\sigma$ detection at 350 $\mu$m; and
\item photometric redshift greater than $z_{\rm ph} \ge 1.5$, as discussed below.
\end{enumerate}
These left a total of 94,825 sources of the H-ATLAS sample, of which 15,611 fall within the \pb survey region (6,080 in the RA12 field and 9,531 in RA23). 
Finally, we create pixelized maps of the galaxy overdensity as 

\begin{equation}
\hat{\delta}_g(\nver) = \frac{n(\nver) - \bar{n}}{\bar{n}},
\end{equation}
where $n$ is the number counts in a $2^\prime$ x $2^\prime$ pixel, $\bar{n}$ is the mean number counts over the RA12 and RA23 footprints separately, and $\nver$ is the unit vector pointing along the line-of-sight. 
The overlapping and usable sky area between H-ATLAS and \pb amounts to approximately 10 deg$^2$.

\subsection{Galaxy Redshift distribution}\label{subsec:sedz}
The knowledge of the galaxies' redshifts, along with their uncertainties, plays a fundamental role. One one hand, it enables to construct pixelized maps of the projected galaxy distribution in the respective redshift bins. On the other hand, it allows to predict the theoretical cross-power spectrum that is ultimately compared to the measured one and through that, carry out the cosmological and astrophysical inference.

Following \citet{Bianchini2015,Bianchini2016}, we estimate the photometric redshift of each source by $\chi^2$-fitting the observed \textit{Herschel} photometric points to a typical high-$z$ SED. Our baseline SED choice is that of SMM J2135-0102, "The Cosmic Eyelash" at $z = 2.3$ \citep{Ivison2010,Swinbank2010}, that has been shown by \citet{Lapi2011} and \citet{Gonzalez-Nuevo2012} to be a good template for $z \gtrsim 1$, with a median value of $\Delta z/(1+z)\equiv (z_{\rm ph} - z_{\rm spec})/(1+z_{\rm spec}) = -0.002$ and a normalized scatter of $\sigma_{\Delta z/(1+z)} = 0.115$. 
The redshift-dust temperature degeneracy affecting the SED fitting becomes worse at lower redshifts, thus we restrict the analysis to $z_{\rm ph} \ge 1.5$.
The robustness of the analysis results with respect to the choice of the assumed SED is tested in Sec.~\ref{sec:check_sed}.

\begin{figure}
	\includegraphics[width=\columnwidth]{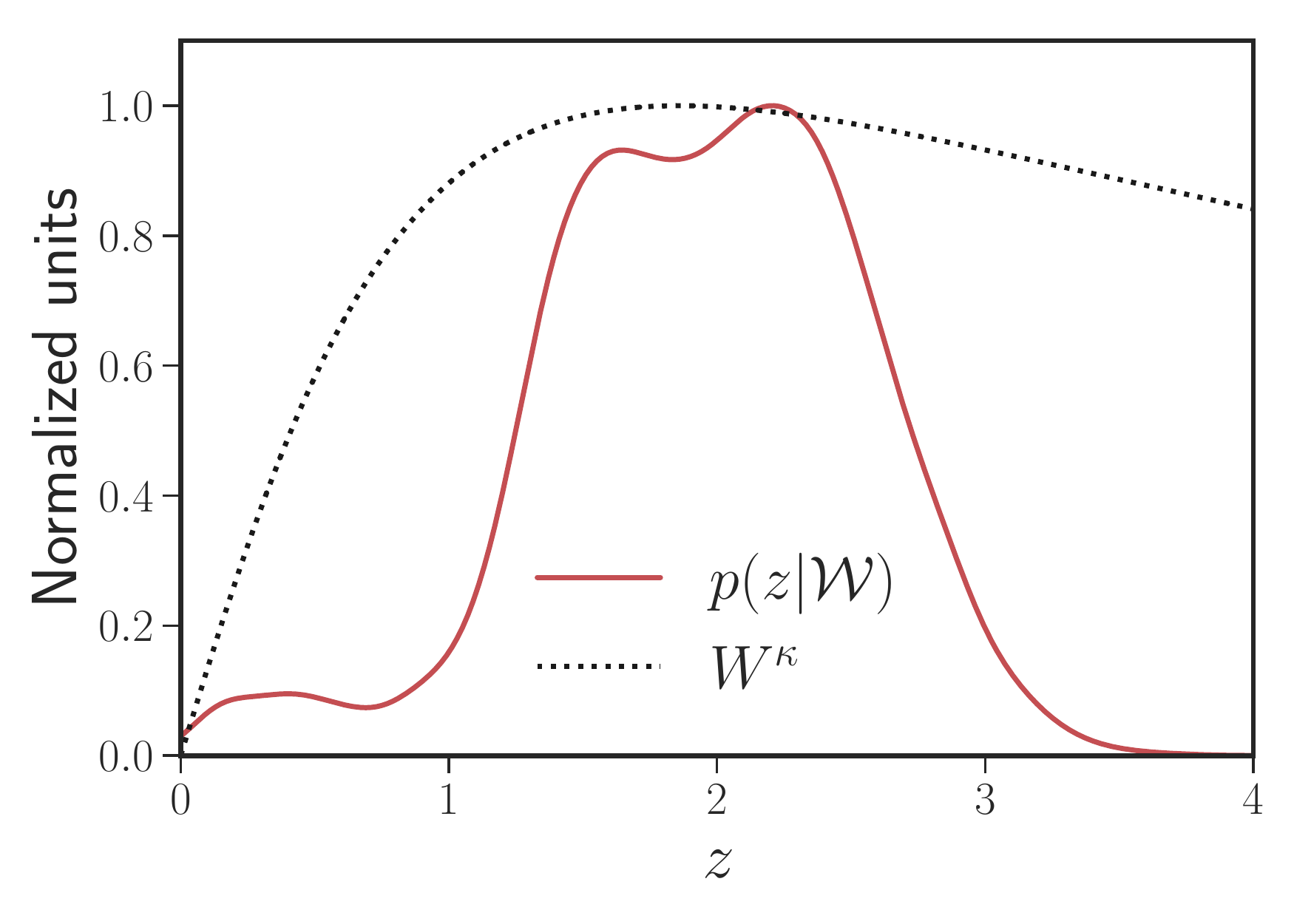}
    \caption{Redshift distribution (solid red line) of the sub-mm sources used in this work allows to probe the peak of the lensing kernel (dotted black line). 
    The CMB lensing kernel $W^{\kappa}$ reflects the relative size of the CMB lensing signal for a fixed mass object at different redshifts; the maximum signal is observed for objects around $1.5 \lesssim z\lesssim 2$.}
        \label{fig:dndz}
\end{figure}

Following \citet{Budavari2003}, we model the redshift distribution $p(z|\mathcal{W})$ of galaxies selected by our window function $\mathcal{W}(z_{\rm ph})$ as:

\begin{equation}
p(z|\mathcal{W}) = p(z) \int dz_{\rm ph} \mathcal{W}(z_{\rm ph})p(z_{\rm ph}|z),
\end{equation}
where $p(z)$ is the fiducial redshift distribution, $\mathcal{W}(z_{\rm ph})$ is 1 for $z_{\rm ph}$ in a selected photo-$z$ interval and 0 otherwise. $p(z_{\rm ph}|z)$ is the probability that a galaxy with a true redshift $z$ has a photometric redshift $z_{\rm ph}$ and is parameterized as a Gaussian distribution with zero mean and scatter $(1 + z) \sigma_{\Delta z/(1+z)}$. The resulting redshift distribution is shown in Fig.~\ref{fig:dndz}. We normalize $p(z|\mathcal{W})$ to unity and finally calculate the redshift distribution $dN/dz$ as $\frac{dN}{dz}= \left(\int dz' p(z'|\mathcal{W})\right)^{-1}p(z|\mathcal{W})$.
We do not account for the effect of catastrophic redshifts failures in the modelling. In fact, when comparing the photo-$z$ estimated with the SMM or \citet{Pearson2013} template (see Sec.~\ref{sec:check_sed} for a robustness check using another SED template) to a subset of sources with known spectroscopic redshift, outliers (defined as those objects for which $|\Delta z/(1+z_{\rm spec})| > 0.3)$ are much less than 10\%, as it can be seen in e.g. Fig. 5 of \citet{ivison16}.  In particular, outliers become more important at redshifts well below $z < 0.5$ (see Fig. 6 of \citet{Pearson2013}), but in our analysis we considered only objects with estimated photo-$z$ larger than $z \ge 1.5$ to mitigate this effect as much as possible.

\subsection{Galaxy Overdensity Simulations}

To generate realizations of the galaxy field comprising of signal and noise with statistical properties that match those of the data, we follow the approach in \citet{Smith2007}. We start by generating a simulated galaxy counts map, where the value at each pixel $p$ is spatially modulated by a Gaussian field $g$ generated from the fiducial galaxy auto-spectrum $C_L^{gg}$. For each pixel, this is accomplished by drawing a number from a Poissonian distribution with mean $\lambda(p) = \bar{n} ( 1 + g(p))$, where $\bar{n}$ is defined in Sec.~\ref{sec:hatlas}. Finally, we convert the galaxy counts map to overdensity as done for the real galaxies.

\section{Theory}
\label{sec:theory}
The observed CMB lensing  and galaxy overdensity fields trace the same
underlying matter fluctuations in different and complementary ways. 
Galaxies are biased signposts of the same dark matter haloes that are lensing the CMB photons. 
Whereas lensing probes the integrated matter distribution along the line-of
sight, galaxy surveys provide a biased sparse sampling of the dark matter field. 
Both the projected CMB lensing convergence $\kappa(\nver)$
and galaxy overdensity $\delta_g(\nver)$ fields along a given line-of-sight $\nver$ can be
expressed as a weighted integral of the 3D dark matter density contrast $\delta$,

\begin{equation}
X(\nver) = \int_0^{z_*} dz\, W^X(z)\delta(\chi(z)\nver,z).
\end{equation}
Here $X=\{\kappa,\delta_g\}$ and the two fields' response to the underlying matter distribution is encoded by the kernels $W^X(z)$, while $\chi(z)$ denotes the comoving distance to redshift $z$. 
In the case of CMB lensing convergence, the kernel is given by

\beq
\label{eqn:wk}
W^{\kappa}(z) = \frac{3\Omega_m}{2c}\frac{H_0^2}{H(z)}(1+z)\chi(z)\frac{\chi_*-\chi(z)}{\chi_*},
\eeq
where $H(z)$ is the Hubble factor at redshift $z$ and $\chi_*$ is the comoving distance to the last scattering surface. $\Omega_m$ and $H_0$ are the present-day values of matter density and Hubble parameter, respectively.

The galaxy overdensity kernel can be written as the sum of two terms, one describing the intrinsic clustering of the sources and one quantifying the so-called magnification bias effect, the apparent clustering of the sources due to the lensing by foreground matter clumps \citep{Turner1980,Moessner1998}:

\begin{align}
\label{eqn:wg}
W^{ \delta_g}(z) &= b(z)\frac{dN}{dz} + \mu(z),\\
\mu(z) &= \frac{3\Omega_{\rm m}}{2c}\frac{H_0^2}{H(z)}(1+z)\chi(z) \\
&\times\int_z^{z_*}dz'\,\left(1-\frac{\chi(z)}{\chi(z')}\right)(\alpha(z')-1)\frac{dN}{dz'}.
\end{align}
In the above equation, we have assumed a linear, local, and deterministic galaxy bias $b(z)$  to relate the galaxy overdensity $\delta_g$ to the matter overdensity $\delta$ \citep{Fry1993}, while $dN/dz$ denotes the unit-normalized redshift distribution of the galaxy sample (we use the red solid curve in Fig.~\ref{fig:dndz}). Note that the magnification bias term is independent on the galaxy bias parameter but depends, in the weak lensing limit, on the slope $\alpha(z)$ of the integrated galaxy number counts above the flux density limit $S_{\rm min}$ of the survey

\begin{equation}
N(>S_{\rm min})\propto S^{-\alpha}.
\end{equation}  
For the high-$z$ galaxies selected in this work, \citet{Gonzalez-Nuevo2014,Bianchini2015} have shown that the magnification bias is substantial. The reason is that the source counts are steep. In fact, the slope of the integrated number counts at the flux limit, as measured from the data at 250 $\mu$m where the main selection is operated, is $\alpha \simeq 3$. 

Given that we are interested in sub-degree and degree angular scales ($\ell \gg 10$), we can safely adopt the so-called Limber approximation \citep{limber} and evaluate the theoretical cross-power spectrum at a given angular multipole $\ell$ as:

\begin{equation}
\label{eqn:limber}
C_{\ell}^{\kappa g} =   \int_0^{z_*} \frac{dz}{c} \frac{H(z)}{\chi^2(z)} W^{\kappa}(z)W^{\delta_g}(z)P_{\rm NL}\left(k=\frac{\ell+\tfrac{1}{2}}{\chi(z)},z\right).
\end{equation}
We compute the non-linear matter power spectrum $P_{\rm NL}(k,z)$ using the \texttt{CAMB}\footnote{\url{https://camb.info/}} code \citep{camb}  with the \texttt{Halofit} prescription of \citet{Takahashi2012}.  As can be seen from Eq.~\ref{eqn:limber}, the cross-power spectrum $C_{\ell}^{\kappa g}$ is sensitive to the parameter combination $b(z)\sigma_8^2(z)$, where $\sigma_8(z)$ measures the amplitude of the (linear) power spectrum on the scale of 8 $h^{-1}$ Mpc at redshift $z$.

Under the assumption that both the CMB lensing potential and the galaxy overdensity fields behave as Gaussian random fields, we can forecast the expected signal-to-noise ($S/N$). 
For the survey specifications discussed above, and an assumed galaxy bias $b = 3$ \citep{Bianchini2015,Amvrosiadis2018}, we forecast an overall $S/N$ for $100 \le L \le 1500$ of 3.4.
This is somewhat smaller than the observed value presented in Sec.~\ref{sec:results}.  As we shall see, the reason is that the inferred galaxy bias value is larger than the one assumed for this S/N forecast.
 
\section{Methods}
\label{sec:methods}
\subsection{Power Spectrum Estimation}
\label{sec:power_spectrum_estimator}
\begin{figure}
	\includegraphics[width=\columnwidth]{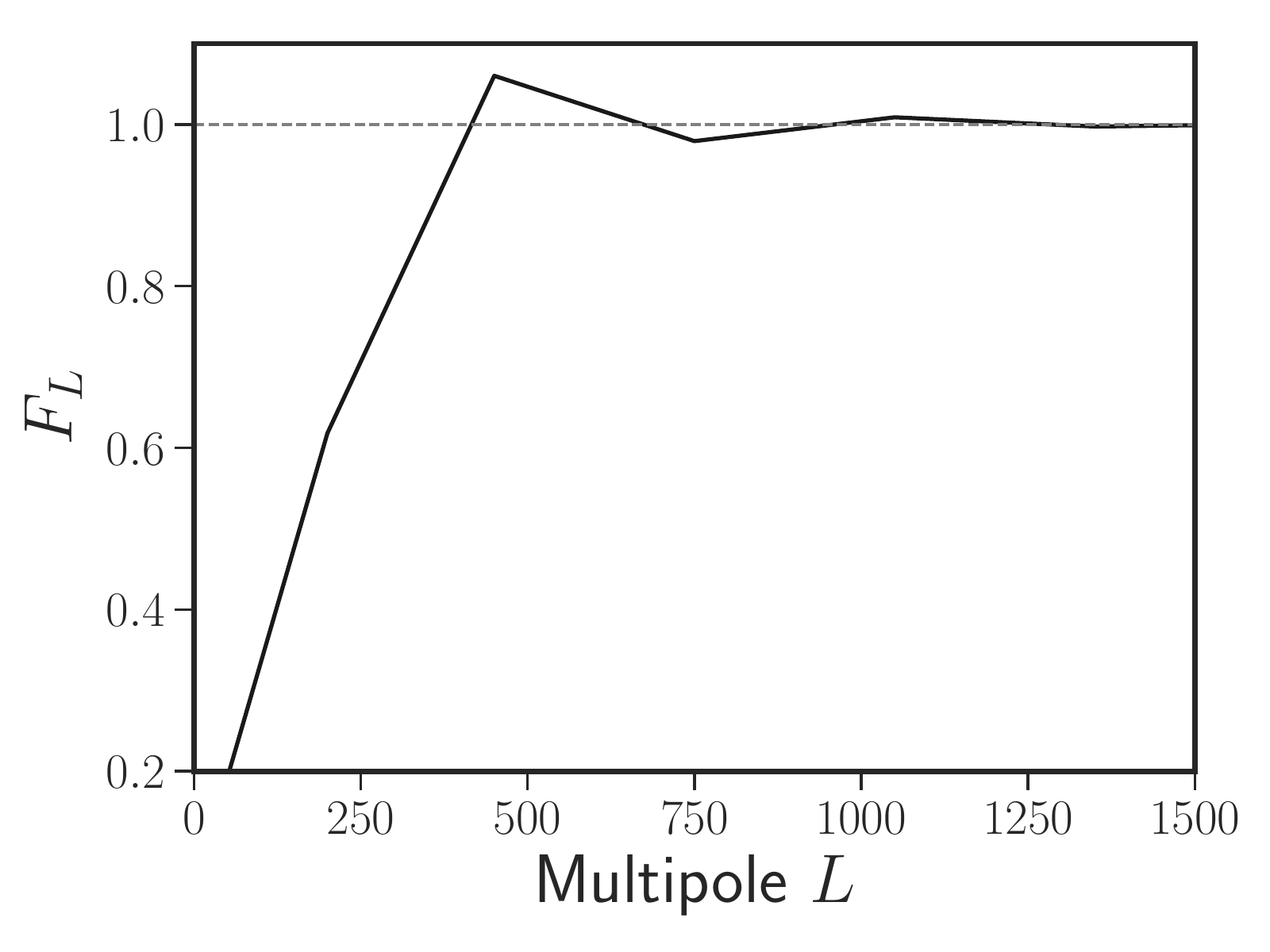}
    \caption{Harmonic counterpart of the real-space filter that we apply to the CMB lensing and galaxy density maps.}
        \label{fig:filter_function}
\end{figure}

We measure the cross-correlation signal between CMB lensing and the spatial galaxy distribution in the Fourier domain. As a first step, we perform a real-space convolution between both the CMB lensing $\hat{\kappa}$ and $\hat{\delta}_g$ data maps and a tapering function, in order to minimize the noise leakage from large scale to small scales \citep[e.g.,][]{Das2009}. 
Even though the convolution is performed in real-space, we use simulations to evaluate the corresponding transfer function $F_L$ that allows us to deconvolve the final cross-power spectrum measurement for its effect following \citet{Hivon2002}. The shape of the filter function in the Fourier domain is shown in Fig.~\ref{fig:filter_function}.

After multiplying the convolved maps by the mask, we calculate the Fourier transforms of the observed fields 
$\hat{\kappa}(\Bell)$ and $\hat{\delta}_g(\Bell)$, and estimate the 1D cross-power spectrum of the windowed maps 
in the flat-sky approximation as 

\begin{equation}
\hat{C}_L^{\kappa g} = f_{\rm sky}^{-1}\left\langle \Re\left[\hat{\kappa}(\Bell) \hat{\delta}_g^*(\Bell)\right] \right\rangle_{\Bell \in L}, 
\end{equation}
where $\Re$ denotes the real part of a complex quantity, the average is over all the pixels in the Fourier plane $\Bell = (\ell_x,\ell_y)$ that fall within the bandpower associated to $L$, and the `` $\hat{}$ " indicates quantities measured from the data. We account for the effect of masking by rescaling the observed power for $f_{\rm sky}$, the effective area calculated as the mean of the squared mask.
Similar to our previous analysis of \citet{PBIPPxHerschel}, the cross-power spectrum is reconstructed in five multipole bins between $100 \le L \le 1500$, probing physical scales between 55 Mpc and 4 Mpc at an effective redshift of $\langle z \rangle \sim 2$. We generate 500 correlated CMB lensing and galaxies overdensity simulations that include both the sky signal and noise to check that the power spectrum estimator correctly recovers the input theory within the measured errors without introducing any spurious correlations.

After extracting the power spectra for each lensing quadratic estimator $\alpha=\{EE,EB\}$ and field $f= $\{RA12,RA23\}, we co-add the individual $\hat{C}_L^{\kappa g,\alpha,f}$ in a single estimate as

\begin{equation}
\hat{C}_L^{\kappa g} = \frac{\sum_{\alpha,f} w_L^{\alpha,f} \hat{C}_L^{\kappa g,\alpha,f}}{\sum_{\alpha,f} w_L^{\alpha,f}}, 
\end{equation}
where the weights $w_L^{\alpha,f}$ for the bandpowers of each field and estimator are given by the inverse of the variance of the bandpowers, i.e. the diagonal component of the covariance matrices. 

The covariance matrices $\mathbb{C}_{LL'}$ are estimated in a Monte Carlo (MC) approach by cross-correlating the observed H-ATLAS density maps with 500 simulated \pb lensing convergence maps. We have checked that cross-correlating the true \pb CMB lensing convergence maps with simulated galaxy density maps with statistical properties that match that of the data yields comparable error bars.

\subsection{Null Tests}
We adopt a blind analysis strategy to mitigate observer bias and increase the robustness of our results. Before unblinding the cross-power spectrum, we thus perform a suite of 15 null tests to test for the presence of systematic effects summarized in \tabref \ref{tab:null-suite}. Within this suite, we define two sub-suites: the \pb suite and the Analysis suite.

\subsubsection{\pb\ Suite}
The \pb suite definition follows PB17 and consists of 12 different data splits sensitive to multiple sources of instrumental systematic contamination, such as contaminations due to the atmosphere or the telescope sidelobes pickup, systematic effects in the telescope beam or detector response, and vibrations due to the telescope motion.  
From the two halves of each \pb suite data splits we reconstruct two CMB convergence maps and then take the difference between them before computing the null cross-power spectrum with the galaxy density.

\subsubsection{Analysis Suite}
The Analysis suite consists of a test aimed at assessing the consistency of the source catalog (galaxy catalog null test), and two other tests targeting the robustness of the analysis pipeline, namely the swap-field and the curl null test. In the galaxy catalog null test, we compute the difference of two galaxy overdensity maps of two random halves of the H-ATLAS catalog and correlate it with the \pb convergence maps. In the swap-field test, we cross-correlate the \pb maps with non-overlapping \textit{Herschel} galaxy density maps, e.g. \pb RA23 with \textit{Herschel} RA12. Finally, for the curl null test, we reconstruct the curl component of the CMB lensing field (which is expected to be zero at linear order \citep{fabbian2018} and without systematic artifacts) and cross-correlate it with the \textit{Herschel}  map on the same sky region.

\begin{table}[]
\caption{Summary of the null tests performed in this work.}
\begin{tabular}{c|c}
\toprule
Null test suite & Null test type\\ 
\midrule
\multirow{12}{*}{\pb}             		      & Dataset first half vs second half             \\ 
                                                                                          & First season vs second season         \\
                                                                                          & Focal plane pixel type\footnote{Focal plane detectors have two different polarization angles orientations.}                       \\
                                                                                          & High vs low elevation observations      \\
                                                                                          & Rising vs setting                     \\
                                                                                          & High gain vs low gain                 \\
                                                                                          & Good vs bad weather                   \\
                                                                                          & Moon distance                         \\
                                                                                          & Sun distance                          \\
                                                                                          & Sun above vs below the horizon        \\
                                                                                          & Left vs right side of the focal plane \\
                                                                                          & Left vs right-going subscans\footnote{Observations are divided into constant elevation scans: each sweep in azimuth is defined as a subscan.}    \\ \midrule
\multirow{3}{*}{Analysis} & Galaxy catalog                  \\
                                                                                          & Curl mode                                 \\
                                                                                          & Swap-field            \\              
\bottomrule
\end{tabular}
\label{tab:null-suite}
\end{table}

\subsubsection{Null Test Statistics}

For each null spectrum band power $L$, we calculate the statistic $\chi^{\rm null}_L \equiv \hat{C}_{L}^{\rm null}/\sigma_L$, where $\sigma_L$ is a MC-based estimate of the standard deviation of the null spectra. In our null test framework, we use both $\chi^{\rm null}_L$ and $(\chi^{\rm null}_L)^2$ as the former is sensitive to systematic biases while the latter is mostly sensitive to outliers. From these two quantities, we compute four statistics for the minimum-variance cross-power spectrum to test for systematic contamination affecting a particular test or bin: i) average $\chi_{\rm null}$ over all tests and bins; ii) most extreme $\chi^2_{\rm null}$ by bin when summing over all null tests; iii) most extreme $\chi^2_{\rm null}$ by null test when summing over all bins; iv) total $\chi^2_{\rm null}$, summed over all tests and bins. For each of these statistics, we calculate the probability-to-exceed (PTE) by comparing the statistic value found for real data with values found in MC simulations. Additionally, we compute a Kolmogorov-Smirnov (KS) test by comparing the distribution of the $\chi^2_{\rm null}$ by test and by bin PTEs to an uniform distribution. 
In order to consider the null tests passed, we require that the two PTEs of the KS tests and the worst PTEs of the four statistics discussed above are larger than 5\% for both the \pb and Analysis suites individually and combined. As can be seen in Tab.~\ref{tab:null_tests}, these requirements are all met. Therefore, we find that systematic effects are well below the statistical detection level.

\begin{center} 
\begin{deluxetable*}{cccc}
\tabletypesize{}
\tablecolumns{4}
\tablewidth{\columnwidth}
\tablecaption{Null test suites summary statistics. \label{tab:null_tests}}
\tablehead{
\colhead{Null test suite} & \colhead{Worst PTE} &  \colhead{KS by bin} & \colhead{KS by test} }
\startdata
All tests & 28\%  & 62\% & 85\%\\
\pb &  28\% & 67\% & 92\% \\
Analysis & 55\% & 97\% & 92\%\\
\enddata
\end{deluxetable*}
\end{center}
%
\section{Results}
\label{sec:results}
\subsection{Cross-power Spectrum}
\begin{figure}
	\includegraphics[width=\columnwidth]{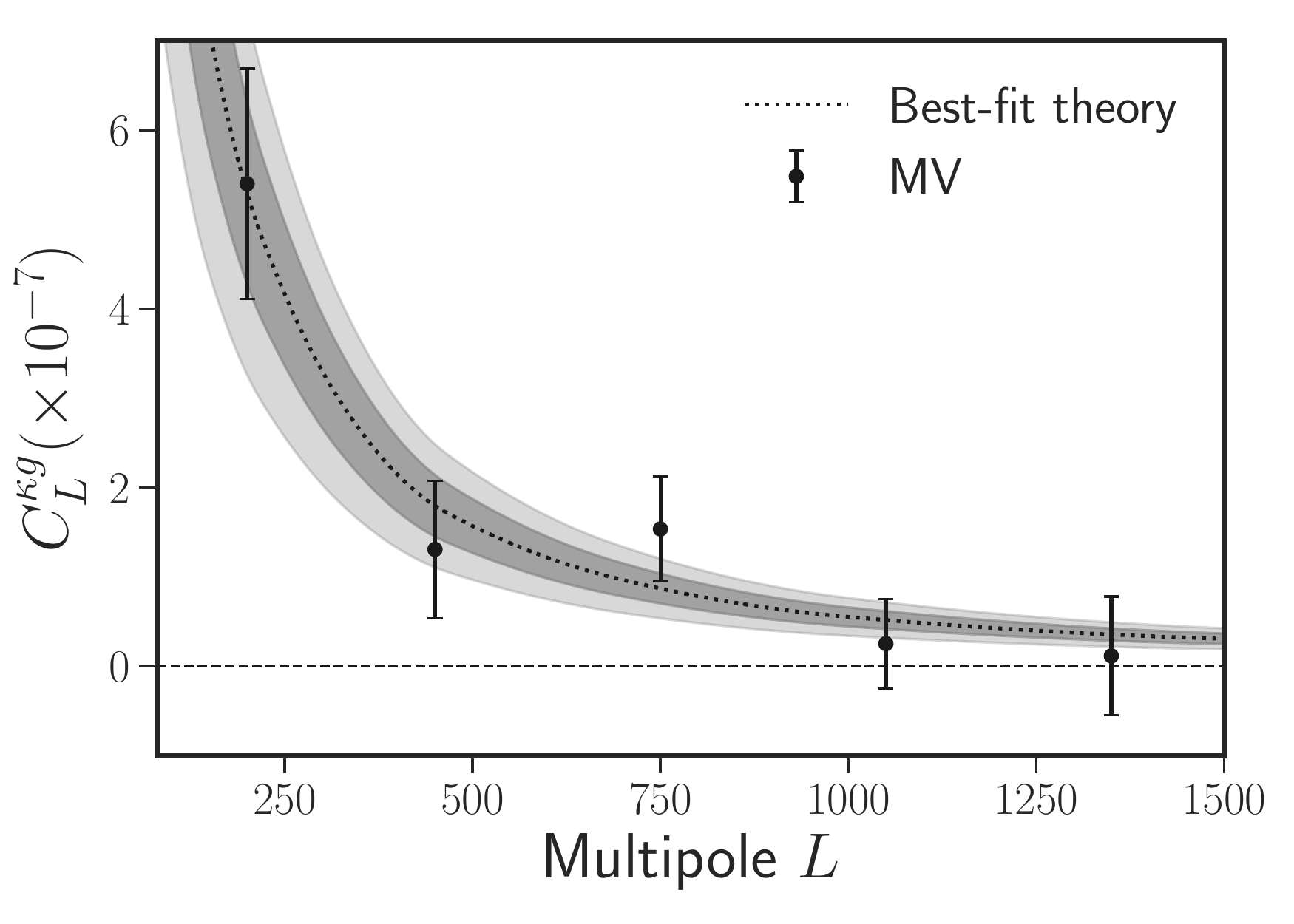}
    \caption{Final cross-power spectrum between \pb CMB lensing and the H-ATLAS galaxy overdensity in RA12 and RA23 patches (black circles). The dotted black line represents the best-fit theoretical model (corresponding to a galaxy bias of $b=\bvalue$), while the dark and light grey shaded regions indicate the 1 and $2\sigma$ uncertainties  respectively.}
    \label{fig:clkg}
\end{figure}

The final cross-power spectrum between the \pb CMB lensing convergence maps and the H-ATLAS galaxy overdensity is shown in Fig.~\ref{fig:clkg}. As mentioned in Sec.~\ref{sec:power_spectrum_estimator}, we calculate the error bars on the band powers by cross-correlating 500 realizations of the CMB lensing field as reconstructed by the \pb pipeline with the real H-ATLAS maps. By doing so, the two maps are uncorrelated, which turns out to be a well-founded assumption since $(C_L^{\kappa\kappa}+N_L^{\kappa\kappa})(C_L^{gg}+N_L^{gg}) \gg (C_L^{\kappa g})^2$ over the relevant scales. More quantitatively, adopting our fiducial cross-correlation model, we have checked that neglecting the cross-power spectrum term leads to an underestimation of the uncertainties of about 14\% for the first bin and less than 5\% for the second bandpower.
We also note that the covariance matrix is dominated by the diagonal elements, with a neighbouring bins correlation of at most $\approx 15\%$. A statistically significant cross-power is detected. We define the null-hypothesis as the absence of correlation between the CMB lensing and the galaxy fields, i.e. $C_L^{\kappa g} = 0$. Then, the chi-square value under this null-hypothesis can be evaluated as $\chi^2_{\rm null} = \sum_{LL'}\hat{C}_L^{\kappa g}\mathbb{C}^{-1}_{LL'}\hat{C}_{L'}^{\kappa g} \simeq 26.1$.

\subsection{Constraints}
As we have seen in Sec.~\ref{sec:theory}, the theoretical cross-power spectrum $C_L^{\kappa g}$ depends on cosmology, for example through the $\Omega_m H_0^2 \sigma_8$ combination and astrophysical parameters, such as the galaxy bias $b$. Here, we fix the underlying cosmology and fit for the linear galaxy bias. For reference, the assumed values of matter density, Hubble constant (in km s$^{-1}$Mpc$^{-1}$), and $\sigma_8$ are $\{\Omega_{\rm m},H_0,\sigma_8\}= \{0.3153, 67.36, 0.8111\}$. The large number of effective independent modes in each band power allows us to assume a Gaussian likelihood as $-2 \ln \mathcal{L}(\hat{C}_L^{\kappa g}| b) \propto \chi^2$, where $\chi^2 = \sum_{LL'}\left[\hat{C}_L^{\kappa g} - C_L^{\kappa g}(b)\right] \mathbb{C}_{LL'}^{-1} \left[\hat{C}_{L'}^{\kappa g} - C_{L'}^{\kappa g}(b)\right]$. The posterior space is then sampled through a Markov chain Monte Carlo (MCMC) method implemented in the publicly available \texttt{emcee} code \citep{emcee}. The resulting best-fit galaxy bias is $b=\bvalue$ with a corresponding $\chi^2_{\rm bf}\simeq 2.5$ for $\nu = 5-1=4$ degrees-of-freedom, or a PTE of about 64\%.\footnote{The central value and the $\pm 1 \sigma$ uncertainties are evaluated as the 50th and 16th/84th percentiles of the posterior distribution respectively.}  The significance is computed as the square-root of the difference between the null-line chi-squared value ($b=0$) and the best-fit theory line, $S/N = \sqrt{\chi^2_{\rm null}-\chi^2_{\rm bf}} \simeq 4.8$.

To give a sense of how an assumption of different cosmological parameters propagates into the inferred constraints on the galaxy bias, we perturb $\sigma_8$ by $\pm 3\%$. The corresponding galaxy biases are found to be $b = 6.22 \pm 1.26$ and $b = 5.49 \pm 1.12$ (negative and positive perturbations respectively). The differences with respect to the baseline galaxy bias constraint are well within the statistical uncertainty.

The modelling of the magnification bias, encoded by the parameter $\alpha$, also affects the inferred galaxy bias value. To quantify its impact, we have obtained constraints on the galaxy bias assuming two different fiducial values of $\alpha$, namely an unrealistic case where there is no magnification bias ($\alpha = 1$) and  $\alpha = 5$. The respective constraints are $b = 9.25^{+1.93}_{-2.04}$ and $b = 4.27 \pm 0.94$. As expected, by boosting the expected $C_{\ell}^{\kappa g}$, a larger value of $\alpha$ corresponds to a lower galaxy bias.

The galaxy bias constraint can be translated into an estimate of the effective mass $M_h$ of the dark matter haloes inhabited by the H-ATLAS galaxies. We assume the bias model provided by \citet{Tinker2010} and relate the scale-independent galaxy bias to the peak height of the linear density field $\nu = \delta_c/\sigma(M)$, where $ \delta_c=1.686$ is the critical threshold for spherical collapse and $\sigma(M)$ is the root mean square density fluctuation for a mass $M$. We infer that, at an effective redshift of $\langle z \rangle \sim 2$,  these sub-mm galaxies  are hosted by haloes of characteristic mass of $\log_{10}(M_h/M_\odot) = \Mhalovalue$.\footnote{We adopt a ratio between the halo mass density and the average matter density of  $\Delta =200$.}

From the observational point of view, several authors have studied the clustering properties of galaxies selected at both short (250-500 $\mu$m) and long (850-1200 $\mu$m) sub-mm wavelengths. 
Numerical simulations have shown that the expected characteristic mass of haloes inhabited by sub-mm sources at $z \sim 2$ is $M_h \sim 10^{13} M_{\odot}$  \citep[e.g.][]{Dave2010,McAlpine2019}. However, a direct comparison between the mass estimates found in different studies is complicated by a number of selection effects that affect the galaxies samples being analyzed.  As a result, the inferred halo mass range spans about 1 dex \citep[e.g.,][]{Casey2014,Cowley2016,Wilkinson2017}. Nonetheless, we attempt to place our measurement in the broader context of similar analyses.

The first thing to notice is that the galaxy bias inferred, or the corresponding effective halo mass, seems to fall in the higher end of the mass spectrum found by previous studies, although the uncertainties are relatively large. 
For example, \citet{Cooray2010} measured the angular correlation function of \textit{Herschel} galaxies at $z \sim 2$ with $S_{250\mu m}>30$ mJy and inferred a bias of $2.9 \pm 0.4$, corresponding to effective halo masses of $(5 \pm 4) \times 10^{12}M_{\odot}$. 
For sub-mm galaxies between $1 \lesssim z \lesssim 3$ detected at 870 $\mu$m with LABOCA, \citet{Hickox2012} have derived a corresponding dark matter halo mass of $\log_{10}(M_h /[M_{\odot}h^{-1}])=12.8^{+0.3}_{-0.5}$, consistent with measurements for optically selected quasi-stellar objects.
Similarly, clustering measurements of the bright sub-mm galaxies detected by SCUBA-2 at 850 $\mu$m  by \citet{Wilkinson2017} suggest that these objects occupy high-mass dark matter halos ($M_h > 10^{13}M_{\odot}$) at redshifts $z > 2.5$.
More recently,  \citet{Amvrosiadis2018} have measured the angular correlation function of the sub-mm H-ATLAS galaxies with flux densities $S_{250\mu m}>$30 mJy within the NGP and GAMA fields, finding that they typically reside in dark matter haloes of mass $\log_{10}(M_h/[M_{\odot} h^{-1}]) = 13.2 \pm 0.1$ across the redshift range $1 < z < 5$.

Finally, we note that similar information can be extracted from the clustering of CIB fluctuations with the caveat that, differently from catalog-based analysis, diffuse CIB includes emission from unresolved galaxies with fainter far-IR luminosities (hence less massive). For example, \citet{Viero2009} analyzed the CIB anisotropies measured at 250 $\mu$m by BLAST and inferred a bias of $3.8\pm 0.6$ or an effective mass of $\log_{10}(M_h/M_{\odot})=12.9\pm 0.3$, while from the angular power spectrum analysis of the CIB fluctuations from \textit{Planck}, \textit{Herschel}, SPT and ACT, \citet{Xia2012} found an effective halo mass $\log_{10}(M_h /M_{\odot})\sim 12.7$ (no errors given) for sub-mm galaxies at $z \sim 2$.

To further test the consistency of our results, we follow \citet{Bianchini2015,Giannantonio2016,Omori2018} and introduce an overall multiplicative bias $A$ that scales the cross-correlation as $\hat{C}_L^{\kappa g} = A C_L^{\kappa g}(b)$. We can interpret $A$ as the lensing amplitude, and  a value of $A$ different from unity can be ascribed to the presence of systematics, to improper modelling of the signal, to a mismatch in the assumed underlying cosmology, and possibly to new physics. Of course there will be a degeneracy between the amplitude $A$ and the galaxy bias, since the cross-power spectrum probes a combination of $bA$. In fact, the aforementioned studies combine the cross-correlation and the galaxy clustering measurements, that scales as $b^2$ albeit at the price of being more prone to systematics, to break such degeneracies. Nonetheless, we adopt the same MCMC approach outlined above and infer a constraint on $bA = \bAvalue$. In light of the above discussion on the bias constraints from literature, one would expect a $bA$ value around 3 $\sim$ 4. This value is approximately 1.5 $\sim 2\sigma$ lower than we measured. 

When examining the cross-spectrum of the two fields separately, we find that the high value is coming from the RA12 field. 
A plausible explanation for the high value of the bias is that the lensing signal in this small ($\sim$\,5\,\sqdeg) field has scattered high due to sample variance.  
The \pb{} lensing map is sample variance limited in the first two multipole bins,  which drive the amplitude constraint. A precise estimate of the significance of the observed scatter in the RA12 cross-spectrum is not straightforward to quantify without, for example, the knowledge of the exact galaxy bias value. Given that the power excess is driven by the first band power around $L\sim 200$, we have performed the following check in order to understand whether the scatter is anomalous or not. We have extracted the CMB lensing-galaxy density cross-spectrum $C_L^{\kappa g}$ between a set of correlated (Gaussian) lensing and galaxy realizations and measured the ratio between the first band power value and the statistical uncertainties, $C_L^{\kappa g}/\Delta C_L^{\kappa g}$. We found that, in about 4\% of the simulations, the first band power lies more 4$\sigma$ (the value found in data) away from the null line. We then conclude that, although large, this fluctuation does not seem to be anomalous.
Finally, we note that an excess of power is also observed in RA12 in cross-correlation with galaxy lensing from the Hyper Suprime-Cam \citep{pbxhsc} even though with larger error bars. We also stress that this pipeline used a completely independent lensing reconstruction pipeline. 

\subsection{Comparison with \textit{Planck}}
Our result can be directly compared to the one of \citet{Bianchini2016}. In that paper, the authors correlated the same H-ATLAS sources catalog adopted here with the publicly available all-sky CMB lensing convergence map from \textit{Planck} \citep{PlanckCollaboration2015}. 
Specifically, the authors exploited the full overlap between the H-ATLAS survey and the \textit{Planck} footprint to reconstruct the cross-power spectrum $C_L^{\kappa g}$ between $100 \le L \le 800$. 

\begin{figure}
	\includegraphics[width=\columnwidth]{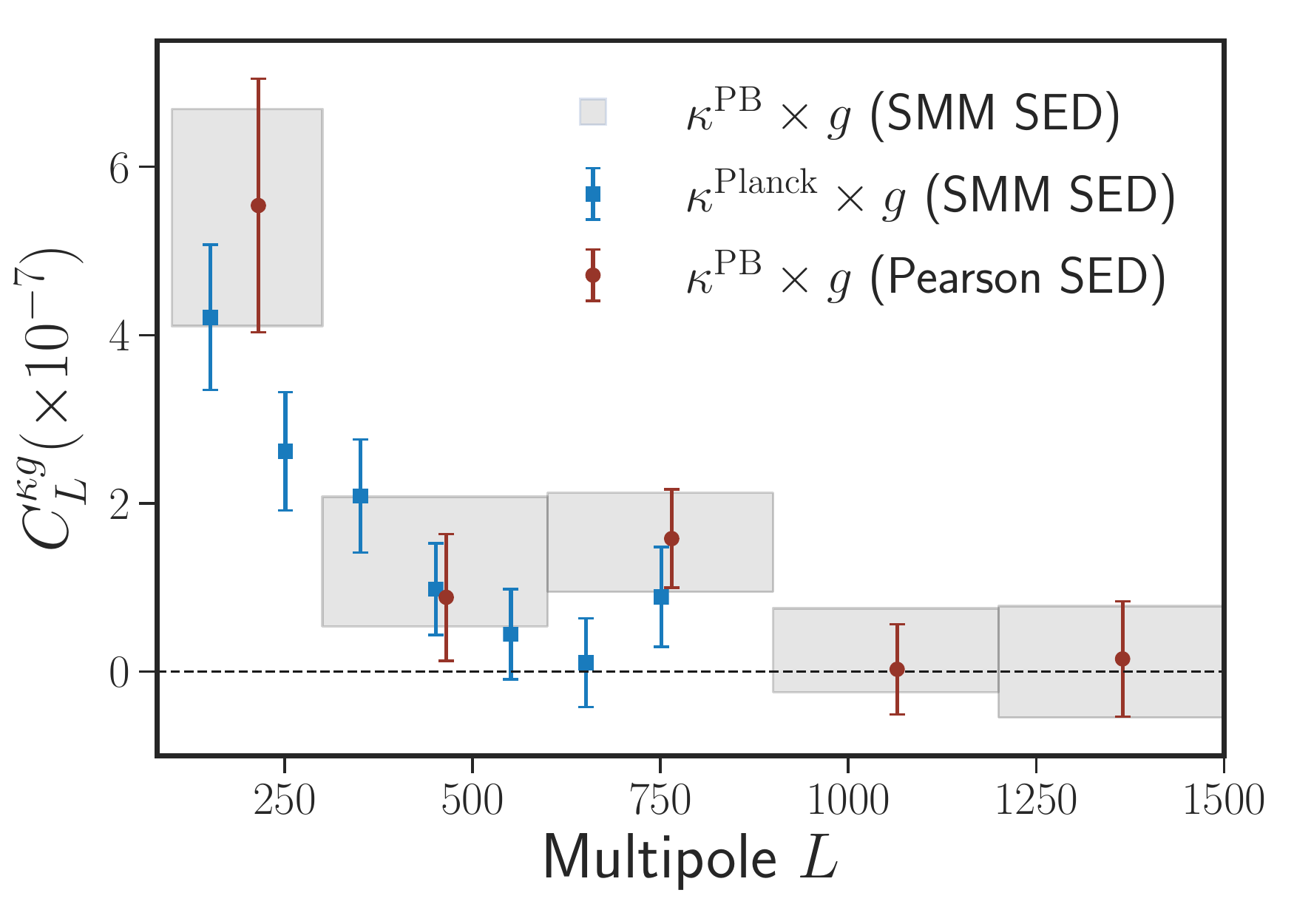}
    \caption{Our baseline measurement of the CMB lensing-galaxy cross-power spectrum over a sky fraction $f_{\rm sky}\approx 3\times 10^{-4}$ (using the SMM template, shown as grey boxes) is in agreement with the $C_L^{\kappa g}$ found adopting a different SED template (\citet{Pearson2013} SED, red circles) and with the one based on \textit{Planck} CMB lensing data over $f_{\rm sky} \approx 0.01$ (band powers from \citet{Bianchini2016}, blue squares).}
    \label{fig:clkg_pearson_planck}
\end{figure}

As it can be seen in Fig.~\ref{fig:clkg_pearson_planck}, the amplitude and shape of cross-spectrum is similar to the one measured with \pb maps over the range of scales where the visual comparison can be performed. It is also interesting to note that, despite the sky coverage being almost 30 times smaller than that of the \textit{Planck} $\times$ H-ATLAS analysis, we still detect a signal only $\sim 4$ times less significant thanks to the high sensitivity of the \pb CMB lensing convergence maps.

The linear galaxy bias inferred by fitting the \textit{Planck} $\times$ H-ATLAS cross-power spectrum $\hat{C}_L^{\kappa g}$ to the theoretical model is $b = \bvaluePlanck$, roughly $2\sigma$ away from the central value found in our analysis. This corresponds to an effective host halo mass of about $\log_{10}(M_h/M_\odot) = \MhalovaluePlanck$. When including the amplitude $A$, the MCMC analysis reveals a constraint of $bA = \bAvaluePlanck$.  

We also stress that our measurement fundamentally differs from the one based on \textit{Planck} data. While \pb lensing convergence maps have been obtained from polarization data only, where the strength of both thermal Sunyaev-Zel'dovich (tSZ) and CIB emissions is greatly reduced because they both are essentially unpolarized, the \textit{Planck} CMB lensing map is dominated by the information provided by CMB temperature (even though the released map is a minimum variance one and includes also polarization data). In principle, residuals of extragalactic foreground emission such as tSZ effect in galaxy clusters or the CIB emission could contaminate the \textit{Planck} CMB lensing map. Because these emissions are correlated between themselves and with the distribution of the LSS \citep{tszXcib}, they could affect the amplitude of the cross-correlation as positive biases. In particular, the H-ATLAS galaxies have fluxes well below the \textit{Planck} detection limits and contribute, at least partially, to any residual CIB emission present in the \textit{Planck} maps. However, semi-analytic estimates indicate that possible induced biases on the temperature reconstruction should not be large at \textit{Planck} sensitivity level \citep{Osborne2014,vanEngelen2014} and systematic checks performed by the \textit{Planck} team found no evidence for such contamination \citep{PlanckLensing2018}. Given the differences that characterize the \pb and \textit{Planck} lensing reconstructions, and since the recovered cross-power spectra shown in Fig.~\ref{fig:clkg_pearson_planck} are in good agreement, it is unlikely that foregrounds represent a major source of contamination in our results. 

\subsection{Effect of the SED Template}
\label{sec:check_sed}
Another aspect worth investigating is the effect of the fiducial SED template on the recovered cross-power spectrum. 
Since the SED plays a crucial role when inferring the photo-$z$s from H-ATLAS photometry, it is important to test the robustness of the results against variations in the assumed SED. 
To this end, we start from the same galaxy catalog introduced in Sec.~\ref{sec:hatlas} and estimate the redshift of each source by fitting the SED template from \citet{Pearson2013}. 
This template consists in a two-temperature modified blackbody synthesized from the \textit{Herschel} PACS and SPIRE flux densities of 40 bright H-ATLAS sources with known spectroscopic redshift (25 of these sources lie at $0.5 < z < 1$ and have optical spec-$z$s while the remaining 15 sources at $z > 1$ have CO spec-$z$s). The  uncertainty in the template is $\Delta z/(1+z)\equiv (z_{\rm ph} - z_{\rm spec})/(1+z_{\rm spec}) = 0.03$ with an r.m.s. of $\sigma_{\Delta z/(1+z)} = 0.26$. 
Using the new redshifts, we find that 5,022 and 7,772 galaxies fall within the RA12 and RA23 patches respectively. 
We re-run the full analysis pipeline with the galaxy overdensity maps constructed from this catalog and extract the cross-power spectrum shown in Fig.~\ref{fig:clkg_pearson_planck} as red circles. 
As can be seen, in this case too we detect a statistically significant signal, rejecting the null hypothesis with a significance of about $\sqrt{\Delta\chi^2}\simeq \signifPearson$, as opposed to \signif\ in the SMM J2135-0102 case. 
From a visual inspection, the cross-spectrum appears consistent with what found in the baseline case, with all the shifts well within the $1\sigma$ uncertainties. 
Band powers errors appear to be slightly larger because of the reduced number of galaxies at $z \ge 1.5$, hence a larger shot-noise. 
A possible explanation is that the inclusion of sources with optical spec-$z$ at $z<1$ in the calibration of the \citet{Pearson2013} template resulted in a redder SED than the average for those sub-mm sources with CO spec-$z$s, which translates into a slight bias towards low-$z$. The galaxy bias analysis reveals a constraint of $b=\bvaluePearson$ when using the \citet{Pearson2013} SED as opposed to $b=\bvalue$ for our baseline case, meaning that the systematic shift $\Delta b \approx 0.39$ is smaller than the statistical uncertainties.

\subsection{H-ATLAS Galaxies Auto-spectrum}
\begin{figure}
	\includegraphics[width=\columnwidth]{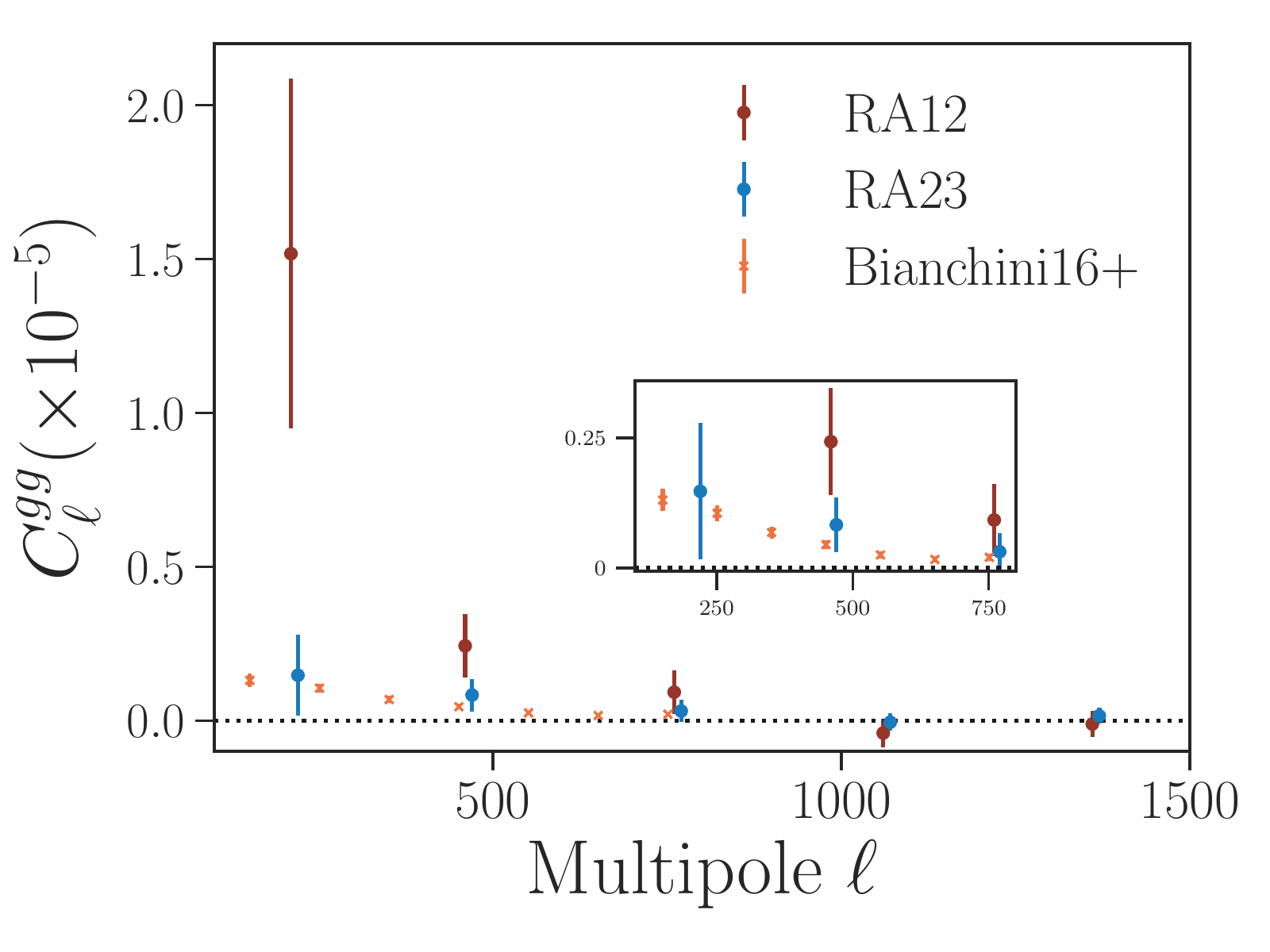}
    \caption{A comparison between the H-ATLAS galaxies auto-power spectrum in the RA12 and RA23 fields (red and blue circles respectively), and the band powers obtained from the full 600 deg$^2$ of the H-ATLAS survey by \citet{Bianchini2016} (blue squares).}
    \label{fig:clgg}
\end{figure}

An informative check to perform is recovering the H-ATLAS galaxies auto-power spectrum in the two fields overlapping with \pb. Performing a thorough analysis of the galaxy auto-power spectrum would require an extensive validation of the measurement that is beyond the scope of the present work, here we naively recover the galaxy auto-spectrum in the RA12 and RA23 fields. An auto-power spectrum analysis of the H-ATLAS galaxies selected with similar criteria as the ones adopted here can be found in \citet{Bianchini2016}. Instead of debiasing the raw galaxy auto-spectra for the shot-noise, we rely on a jackknifing approach. We first randomly split the galaxy catalog in two and create two galaxy overdensity maps, $\delta_g^{(1)}$ and $\delta_g^{(2)}$. From these, we form a pair of half-sum and half-difference maps, $\delta_g^{\pm}=(\delta_g^{(1)}\pm \delta_g^{(2)})/2$. The former map will contain both signal and noise, while the latter will be noise-only. Then we extract their auto-power spectra and evaluate the total galaxy auto-power spectrum as the difference of the half-sum and half-difference overdensity map, $\hat{C}_{\ell}^{gg} = \hat{C}^{+}_{\ell}-\hat{C}^{-}_{\ell}$. The resulting galaxy auto-spectra in RA12 and RA23 are shown as the red and blue points in Fig.~\ref{fig:clgg}. The error bars in the galaxy auto-power spectrum shown in the plot are estimated in the Gaussian approximation using the measured auto-spectra. For comparison, we also include the auto-power spectrum (over the full H-ATLAS fields that cover about 600 deg$^2$) presented in \citet{Bianchini2016}. Even though the uncertainties on the individual RA12 and RA23 fields are large, the power observed in RA23 seems comparable to the full H-ATLAS one, while an excess of clustering seems to be present in RA12. 
%

\section{Conclusions}
\label{sec:conclusions}
In this paper, we have measured the cross-correlation signal between the CMB lensing convergence maps reconstructed from \pb polarization maps and the spatial distribution of the high-$z$ sub-mm galaxies detected by the \textit{Herschel} satellite. Despite the small size of the overlapping patches, the depth of \pb maps together with the redshift extent of the H-ATLAS sources optimally matched to the CMB lensing kernel, have enabled the detection of the cross-power spectrum at a significance of \signif\,$\sigma$. This measurement probes  large-scale structure at an effective redshift $z \sim 2$. 

The cross-correlation power depends on the product  $bA$, with a preferred value of $bA = \bAvalue$. While this is  approximately 2\,$\sigma$ above the expected value of $3\sim4$, we hesitate to interpret this as a tension given the limited statistical evidence. 
The high value is plausibly explained by lensing sample variance over the $\sim$10\,\sqdeg{} of sky. 

We use the galaxy bias information to infer the effective mass of the haloes hosting the H-ATLAS sub-mm sources at a redshift of $z \sim 2$, finding $\log_{10}(M_h/M_\odot) = \Mhalovalue$. This value falls at the high end of the mass spectrum found by previous studies \citep[e.g.][]{Amvrosiadis2018}.

A suite of null tests has been performed to demonstrate that the instrumental systematics are below the statistical detection level. In particular, we stress that lensing reconstructions based on CMB polarization maps, like the one presented in this paper, are less contaminated by galactic and extragalactic foregrounds, providing a clearer view of the projected matter distribution along the line-of-sight. 
Furthermore, the robustness of the results is corroborated by the good agreement between our cross-correlation power measurement and the one based on \textit{Planck} CMB lensing \citep{Bianchini2016}.

Cross-correlations between CMB lensing and LSS tracers, and multi-pronged approaches in general, are becoming a standard tool in cosmological analysis.
In the upcoming years, with the advent of new generation experiments such as the \sa\ \citep{simonsarray} and the Simons Observatory \citep{simonsobs} that will have similar depths over much larger areas of the sky, the full potential of cross-correlation measurements will be unleashed and provide deeper insights on cosmological issues, such as the nature of dark matter, dark energy, and neutrinos, as well as on galaxy formation and evolution.  

\begin{acknowledgements}
The POLARBEAR project is funded by the National Science Foundation under Grants No. AST- 0618398 and No. AST-1212230.
The  James  Ax  Observatory  operates  in  the  Parque
Astron\'omico Atacama in Northern Chile under the auspices of the Comisi\'on Nacional de Investigaci\'on Cient\'ifica
y Tecnol\'ogica de Chile (CONICYT). 
Work at the University of Melbourne is supported by an Australian Research Council's Future Fellowship (FT150100074).
This research was supported in part by the RADIOFOREGROUNDS project, funded by the European Commission€™s H2020 Research Infrastructures under the Grant Agreement 687312, and by the ASI-COSMOS Network of the Italian Space Agency (cosmosnet.it). We also acknowledge support by the INDARK INFN Initiative. 
GF acknowledges the support of the CNES postdoctoral fellowship.
The research leading to these results has received funding from the European Research Council under the European Union's Seventh Framework Programme (FP/2007-2013) / ERC Grant Agreement No. [616170]. 
This research used resources of the National Energy Research Scientific Computing Center (NERSC), a U.S. Department of Energy Office of Science User Facility operated under Contract No. DE-AC02-05CH11231. 
MA acknowledges support from CONICYT UC Berkeley-Chile Seed Grant (CLAS fund) Number 77047, Fondecyt project 1130777 and 1171811, DFI postgraduate scholarship program and DFI Postgraduate Competitive Fund for Support in the Attendance to Scientific Events. 
BK acknowledges support from the Ax Center for Experimental Cosmology. 
YC acknowledges the World Premier International Research Center Initiative (WPI), MEXT, Japan and support from the JSPS KAKENHI grant Nos. 18K13558, 18H04347.
FM acknowledges the support by the World Premier International Research Center Initiative (WPI), MEXT, Japan and JSPS fellowship (Grant number JP17F17025). 
AK acknowledges the support by JSPS Leading Initiative for Excellent Young Researchers (LEADER) and by the JSPS KAKENHI Grant Number JP16K21744.  
Work at LBNL is supported in part by the U.S. Department of Energy, Office of Science, Office of High Energy Physics, under contract No. DE-AC02-05CH11231. 
This work was supported by MEXT KAKENHI grant Nos. 21111002 and JP15H05891, JSPS KAKENHI grant Nos. JP26220709, JP26800125, and JP18H05539, and the JSPS Core-toCore Program.
The APC group acknowledges travel support from Labex Univearths.
We acknowledge the use of many python packages: \sc{IPython} \citep{ipython}, \sc{matplotlib} \citep{Hunter:2007}, \sc{scipy} \citep{scipy}, \sc{SurveyVisualizer}\footnote{\url{https://github.com/Samreay/SurveyVisualiser}}.

\end{acknowledgements}



\bibliographystyle{aasjournal}
\bibliography{references}

\end{document}